\documentclass[a4paper,12pt]{article}

\usepackage{epsfig,graphicx,amsmath,amssymb}
\usepackage{url}
\usepackage{subcaption}
\usepackage[percent]{overpic}
\usepackage{siunitx}
\usepackage{units}
\usepackage{hyperref}
\newcommand{\amu}{a_{\mu}}

\newcommand{\epm}{e^+e^-}

\usepackage{relsize}
\RequirePackage{xspace}
\newcommand{\gev}{\ensuremath{\mathrm{\,Ge\kern -0.1em V}}\xspace}
\def\babar{\mbox{\slshape B\kern-0.1em{\smaller A}\kern-0.1em
    B\kern-0.1em{\smaller A\kern-0.2em R}}}

\def\pipi  {\ensuremath{\pi^+\pi^-}\xspace}
\def\en         {\ensuremath{e^-}\xspace}
\def\ep         {\ensuremath{e^+}\xspace}
\def\Y#1S{\ensuremath{\Upsilon{(#1S)}}\xspace}
\def\FourS {\Y4S}
\def\psiprpr  {\ensuremath{\psi(3770)}\xspace}

\begin{document}

\thispagestyle{empty}

$\phantom{.}$

\begin{flushright}
{\sf  MITP/15-054 \\
  } 
\end{flushright}

\hfill

\begin{center}
{\Large {\bf Mini-Proceedings, 17th meeting of the Working Group on Radiative Corrections and MC Generators for Low Energies} \\
\vspace{0.75cm}}

\vspace{1cm}

{\large 20$^{\mathrm{th}}$ - 21$^{\mathrm{st}}$ April, Laboratori Nazionali di Frascati, Italy}

\vspace{2cm}

{\it Editors}\\
Henryk Czy\.z (Katowice), Pere Masjuan (Mainz), Graziano Venanzoni (Frascati)
\vspace{2.5cm}

ABSTRACT

\end{center}

\vspace{0.3cm}

\noindent
The mini-proceedings of the 17$^{\mathrm{th}}$ Meeting of the "Working Group on Radiative Corrections and MonteCarlo Generators for Low Energies" held in Frascati, 20$^{\mathrm{th}}$ - 21$^{\mathrm{st}}$ April, are presented. These meetings, started in 2006, have as aim to bring together experimentalists and theoreticians working in the fields of meson transition form factors, hadronic contributions to the anomalous magnetic moment of the leptons, and the effective fine structure constant. The development of MonteCarlo generators and Radiative Corrections for precision $e^+e^-$ and $\tau$-lepton physics are also covered, with emphasis on meson production. Heavy quark masses were covered as well in this edition.

\medskip\noindent
The web page of the conference:
\begin{center}
\url{https://agenda.infn.it/conferenceDisplay.py?ovw=True&confId=9287} 
\end{center}
\noindent
contains the presentations.
\vspace{0.5cm}

\noindent
We acknowledge the support and hospitality of the Laboratori Nazionali di Frascati.

\newpage

{$\phantom{=}$}

\vspace{0.5cm}

\tableofcontents

\newpage

\section{Introduction to the $17^{th}$ Radio MontecarLow Working Group meeting}

\addtocontents{toc}{\hspace{1cm}{\sl H.~Czy\.z and G.~Venanzoni}\par}

\vspace{5mm}

\noindent
 H.~Czy\.z$^1$ and G.~Venanzoni$^2$

\vspace{5mm}

\noindent
$^1$Institute of Physics, University of Silesia, 40007 Katowice, Poland\\
$^2$Laboratori Nazionali di Frascati dellÕINFN, 00044 Frascati, Italy

\vspace{5mm}
 
The importance of continuous and close collaboration between the experimental and theoretical groups is crucial in the quest for
precision in hadronic physics. This is the reason why the Working Group on ``Radiative Corrections and Monte Carlo Generators for Low Energies'' (Radio MonteCarLow)  was formed a few years ago bringing together experts (theorists and experimentalists) working in the field of low-energy $e^+e^-$ physics and partly also the $\tau$ community.
Its main motivation was to understand the status and the precision of the Monte Carlo generators (MC) used to analyze the hadronic cross section measurements obtained as well with energy scans as with radiative return, to determine luminosities, and whatever possible to perform tuned comparisons, {\it i.e.} comparisons of MC generators with a common set of input parameters and experimental cuts. This  main effort was summarized in a report published in 2010~\cite{Actis:2010gg}.
During the years the WG structure has been enriched of more physics items and now it includes seven subgroups: Luminosity, R-measurement, ISR, Hadronic VP $g-2$ and Delta alpha, gamma-gamma physics, FSR models, tau decays. 

During the workshop the last achievements of each subgroups have been presented.
The present accuracy and the future prospects of MC generators for  $e^+e^-$ into leptonic, $\gamma\gamma$,  hadronic and tau final states have been reviewed. 
Recent proposals on the hadronic LO contribution to the $g-2$ of the muon in the space-like region and on a dispersive formalism for $\gamma^*\gamma\to \pi\pi$ have been discussed. A new method for extracting heavy quark masses from BES and CELLO data was as well discussed.
New results from CMD3 and BESIII experiments have been presented, together with new Monte Carlo generator developments.

The present workshop, being its 17$^{th}$ edition, was held from the 20$^{th}$ to the 21$^{st}$ April 2015, at the Laboratori Nazionali di Frascati dellÕINFN, Italy.

Webpage of the conference is 
\begin{center}
\url{https://agenda.infn.it/conferenceDisplay.py?ovw=True&confId=9287} 
\end{center}
\noindent
where detailed program and talks can be found.

All the information on the WG can be found at the web page:
\begin{center}
\url{http://www.lnf.infn.it/wg/sighad/} 
\end{center}

\newpage

\section{Summaries of the talks}

\subsection{Recent Results from VEPP-2000: Data and Generators}
\addtocontents{toc}{\hspace{2cm}{\sl S.~Eidelman}\par}

\vspace{5mm}

S.~Eidelman

\vspace{5mm}

\noindent
Budker Institute of Nuclear Physics SB RAS, Novosibirsk 630090, Russia,\\
Novosibirsk State University,  Novosibirsk 630090, Russia \\
\vspace{5mm}

Two detectors at the VEPP-2000 $e^+e^-$ collider in Novosibirsk, CMD-3
and SND, continued data processing of an integrated luminosity of 
$\sim$ 60 fb$^{-1}$ collected by each in 2011 --- 2013 in the center-of-mass  
(c.m.) energy range 320 --- 2000 MeV. Lately analysis has been mainly focused
on the c.m. energies above the $\phi$ meson. In particular, both groups
made an attempt to improve the precision of various cross sections 
important for the muon $g-2$ problem~\cite{ACHASOV:2014nra}.  

Measurements of an integrated luminosity at CMD-3 use two processes,
$e^+e^- \to e^+e^-$ and $e^+e^- \to \gamma\gamma$, allowing a precision of
$\sim 1\%$~\cite{Akhmetshin:2014lea}. At SND, events of large-angle Bhabha
scattering are used to determine an integrated luminosity with a 
systematic accuracy of 2\%~\cite{Aulchenko:2014vkn}.

Both detectors continued analysis of the collected data samples and reported
measurements of cross sections for various processes with pions and $\eta$
mesons. SND published final results on 
$e^+e^- \to \pi^+\pi^-\pi^0$~\cite{snd3pi15} and  
$e^+e^- \to \pi^+\pi^-\eta$~\cite{Aulchenko:2014vkn}. CMD-3 is close to
obtaining final values of $\sigma(2\pi^+2\pi^-2\pi^0)$~\cite{Lukin:2015vsa}.
Analysis of dynamics should be performed simultaneously with a study of
the $3\pi^+3\pi^-$ final state, which cross section had been reported 
before~\cite{Akhmetshin:2013xc}. 

CMD-3 continued studies of various processes with kaons in the final state
using good $K/\pi$ separation based on measuring $dE/dx$ in the drift
chamber. In addition to the process $e^+e^- \to K^+K^-$~\cite{Kozyrev:2015usa},
they reported  close to final results on the cross section and 
dynamics of the $K^+K^-\pi^+\pi^-$ and $K^+K^-\eta$ final states. 
They have also presented their first results on the process 
$e^+e^- \to K^+K^-\pi^0$. 

SND has already published their measurements of the processes with
only neutral particles in the final state -- 
$e^+e^- \to \pi^0\pi^0\gamma$~\cite{Achasov:2013btb}
and $e^+e^- \to \eta\gamma$~\cite{Achasov:2013eli}.

A study of the nucleon form factors near threshold was continued. SND
significantly improved the precision of 
$\sigma(e^+e^- \to n\bar{n})$~\cite{Achasov:2014ncd} compared to the
previous results from FENICE~\cite{Antonelli:1998fv}. CMD-3 measured the
cross section of the process $e^+e^- \to p\bar{p}$
and made an attempt to extract the ratio of the electric and magnetic 
form factors based on the angular distribution of the final 
nucleons~\cite{AKHMETSHIN:2014qta}. 

Both detectors used an original method of Ref.~\cite{cplus} to measure
the partial width of a strongly suppressed $\eta^\prime \to e^+e^-$ decay 
using the inverse process. CMD-3 reported an upper limit of 
$\Gamma(\eta^\prime \to e^+e^-) < 0.0024$ eV at 90\% CL 
based on 2.69 pb$^{-1}$ and one mode of $\eta^\prime$ 
decay~\cite{Akhmetshin:2014hxv}. SND used 2.9 pb$^{-1}$ and five modes of 
$\eta^\prime$ decay to improve it to $< 0.0020$ eV. Finally, they combine
the data samples of CMD-3 and SND to find  
$\Gamma(\eta^\prime \to e^+e^-) < 0.0011$ eV 
at 90\% CL~\cite{Achasov:2015mek} which is still about two orders of
magnitude below the unitary bound. SND has also performed a feasibility
study for a search for $\eta \to e^+e^-$ via $e^+e^- \to \eta$ and concluded
that the only promising decay mode for that is $\eta \to 3\pi^0$. A
dedicated two-week run with the luminosity expected at the c.m.energy 
around the $\eta$ meson mass will allow to improve the existing 
limit~\cite{Agakishiev:2013fwl}. 

A need of high-precision measurements of hadronic cross sections 
and studies of rich resonance dynamics demand Monte Carlo (MC) generators,
which properly take into account interference effects and all symmetries.
Progress in development of the generic MC generator of $e^+e^- \to$ hadrons 
used by CMD-3 for background studies was described~\cite{Czyz:2013sga}.
A generator for production of three pseudoscalars in $e^+e^-$ annihilation
found further applications~\cite{vanderBij:2014mxa}. Intensively discussed
was an important question about radiative corrections included in these and 
similar MC generators.

\newpage

\subsection{Improving the luminosity measurement at BESIII\\
 using the Bhabha event generator Babayaga@NLO}
\addtocontents{toc}{\hspace{2cm}{\sl A. Hafner}\par}

\vspace{5mm}

A. Hafner

\vspace{5mm}
\noindent
Institut f\"ur Kernphysik, Johannes Gutenberg Universit\"at Mainz, \\
J.-J.-Becher-Weg 45, 55099 Mainz, Deutschland.\\
\vspace{5mm}

There is a long standing discrepancy between the experimental measurement of the anomalous magnetic moment of the muon $\amu^{exp}$~\cite{bnl} and its theoretical prediction within the Standard Model of particle physics. This discrepancy amounts to 3-4 standard deviations~\cite{davier,teubner}. New experimental measurements~\cite{fermilab,jparc} will reduce the experimental uncertainty by approximately a factor of 4 within this decade.
The dominant uncertainty on the theoretical prediction of the anomalous magnetic moment of the muon $\amu^{theo}$ stems from the contribution of hadronic Vacuum Polarization (VP), $a_\mu^{VP}=(692.3 \pm 4.2)\cdot 10^{-10}$~\cite{davier}. From causality and analyticity of the VP amplitude a dispersion relation for the VP contribution to $\amu^{theo}$ can be derived~\cite{brodsky}. This relation requires the inclusive hadronic cross section as input. The largest weight is given to low energy contributions. Therefore, the reaction $\sigma(\ep\en\to\pipi)$ contributes with approximately $75\%$.\\

The standard experimental approach is to measure the required hadronic cross sections exclusively at $e^+e^-$ energy scan experiments. Since the last decade, the method of Initial State Radiation (ISR) is used as an alternative approach to measure cross sections of exclusive final states at high luminosity flavor factories, running at a fixed center-of-mass energy. The emitance of a high energy photon from initial state opens the window to low energy hadron physics. KLOE, running on the $\phi$ resonance, measured the $\ep\en\to\pipi$ final state~\cite{kloe:all} with a precision of better than $1\%$ in the peak region. \babar, running on the \FourS resonance, has an extensive ISR-scan program with various final states up to six hadrons from energy threshold up  to 4.5\gev~\cite{babar:data2}. The \babar\ measurement with a precision of $0.5\%$ of the $\pipi$ final state shows a discrepancy in and above the $\rho$ region of up to 2-3 standard deviations to the KLOE measurement. Due to this difference, the resulting uncertainty for $\amu^{theo}$ is similar to the uncertainties of the individual measurements.\\

Energy scan measurements of the $\pipi$ cross section by CMD-3 and SND experiments are expected in the near future with an aimed uncertainty of below $1\%$ and $0.5\%$ respectively. In addition, a new ISR measurement at BESIII, running in the charmonium region and below, is currently performed. The data is statistically and systematically competitive to \babar\ and KLOE. In a first step, the aim is a precision of $1\%$ in the $\rho$-resonance region using a data sample at the \psiprpr.\\ 

The dominating source of systematic uncertainty of this cross section at BESIII stems from the uncertainty of the luminosity measurement, which at present by itself amounts to $1\%$~\cite{psiprprp}. The precision was achieved by measuring Bhabha events and using theoretical input of event generators. Thus, in this measurement, the main contributions to the uncertainty stem from the uncertainty due to the event generator Babayaga.3.5~\cite{baba35} of $0.5\%$ and a very conservative approach for the uncertainty due to the polar angular requirement of $0.7\%$.
The uncertainty of the event generator can be reduced to $0.1\%$ by using the successor of the event generator, Babayaga@NLO~\cite{babanlo}. First studies of the polar angular distribution clearly indicate, that the uncertainty due to the corresponding requirement can also be significantly reduced.
Effects due to VP uncertainties on the Bhabha cross section, and thus the luminosity measurement, have been taken into account and are shown to be negligible for the rather broad \psiprpr resonance.\\

With Babayaga@NLO and the additional systematic studies, the uncertainty for the luminosity will be significantly reduced, allowing the cross section measurement $\sigma(\ep\en\to\pipi)$ at BESIII to reach a precision of $1\%$ and below.\\

I want to thank the organization committee of the $17^{th}$ Radio Monte Carlo working group meeting for the warm hospitality in Frascati and the participants for the fruitful discussions.

\newpage

\subsection{Measurement of the $e^+e^-\rightarrow\pi^+\pi^-$ cross section \\ using initial state radiation 
at BESIII }
\addtocontents{toc}{\hspace{2cm}{\sl C.F.~Redmer}\par}

\vspace{5mm}

\underline{C. F. Redmer}, B. Kloss, and A. Denig

\vspace{5mm}
\noindent
Institut f\"ur Kernphysik, Johannes Gutenberg Universit\"at Mainz, \\
J.-J.-Becher-Weg 45, 55099 Mainz, Deutschland.\\
\vspace{5mm}

Precise measurements of hadronic cross sections at $e^+e^-$ colliders are an important input to the Standard Model 
prediction of the muon anomaly $\amu$. Currently, the accuracy of this prediction is entirely limited by the 
understanding of the hadronic contributions to $\amu$~\cite{JegNyf}. The largest of these contributions stems from the 
hadronic vacuum polarization contribution. It is handled within a dispersive framework, which requires 
$\sigma(e^+e^-\rightarrow \rm hadrons)$ as experimental input~\cite{davier}. The cross section $\sigma_{\pi\pi} = 
\sigma(e^+e^-\rightarrow\pi^+\pi^-)$ contributes more than 70\% to this dispersion relation and is, hence, by far the 
most important exclusive hadronic channel in the endeavor of improving the knowledge on the hadronic contributions to 
$\amu$.

To date, the two most accurate measurements of $\sigma_{\pi\pi}$ have been obtained by the KLOE~\cite{kloe:data}, and 
the BABAR~\cite{babar:data} collaborations. Both experiments claim a precision of better than 1\% in the energy range 
below 1~GeV, however, a discrepancy of approximately 3\% on the peak region of the $\rho(770)$ resonance is observed. 
The discrepancy is even increasing towards higher energies and has a large impact on the SM prediction of 
$\amu$~\cite{davier}. Thus, an independent measurement of competitive accuracy, i.e. in the order of 1\%, is required 
to settle the issue.

This measurement has been performed at the BESIII experiment~\cite{BES3}, operated at the symmetric $e^+e^-$ collider 
BEPCII in Beijing, China. Based on 2.9~fb$^{-1}$ of data~\cite{lumi}, taken at a center-of-mass energy $\sqrt{s}$ = 
3.773~GeV, the method of initial state radiation (ISR) has been exploited, measuring events of the type $e^+e^- 
\rightarrow \pi^+\pi^-\gamma$. In this way, the mass range between 600 and 900 MeV/c$^2$, which corresponds to the 
important $\rho$ peak region, was studied.

The production of muon pairs constitutes the main background contribution in this analysis. To suppress events of the 
kind $e^+e^-\rightarrow \mu^+\mu^-\gamma$, an artificial neural network~\cite{TMVA} has been trained and tested. Monte 
Carlo (MC) simulations of the signal and background final states, $\pi^+\pi^-\gamma$ and $\mu^+\mu^-\gamma$, have been 
used as input. The Phokhara generator~\cite{phok} has been used to generate signal and background samples. Possible 
discrepancies between data and MC simulation due to imperfections in the detector simulation, have to be considered. To 
accomplish a high precision analysis, track-based data-MC correction factors were obtained, comparing the analysis of
nearly background free pion and muon samples in data with MC simulations. 

Two independent normalization methods can be used to extract $\sigma_{\pi\pi}$. On the one hand, the efficiency 
corrected number of $\pi^+\pi^-\gamma$ events $N_{\pi\pi\gamma}$ can be normalized to the luminosity and the radiator 
function~\cite{radi}. On the other hand, the R ratio can be determined, i.e. $N_{\pi\pi}$ is normalized to the number 
of $\mu^+\mu^-­‐\gamma$ events. Both methods have been used in the analysis, and agree within the errors. However, the 
the final result has been obtain based on the first method, since the second one is limited by the $\mu^+\mu^-\gamma$ 
statistics.

The extracted cross section can be used to calculate the two-pion contribution to $\amu$. The preliminary result in the 
mass range between 600 and 900 MeV is found to be $\amu^{\pi\pi,LO} = 374.4 \pm 2.6_{stat} \pm 4.9_{sys}$. Compared to 
the values published by the BaBar and KLOE collaborations for the same energy range, the BESIII result almost coincides 
with their average value and agrees with both results within errors. The preliminary estimated systematic uncertainty 
of 1.3\% of the BESIII measurement is dominated by the uncertainty of the luminosity measurement of 1\%~\cite{lumi}. 
We are confident to reduce this uncertainty by thoroughly re-evaluating the luminosity measurement~\cite{andi}. 
Nevertheless, this result confirms the discrepancy in $\amu$ between SM and experiment on the level of 3 to 4 standard 
deviations.

\newpage

\subsection{Primary Monte-Carlo generator of the process $e^+e^-\to f_0(1370)\rho(770)$ for the CMD-3 experiment}
\addtocontents{toc}{\hspace{2cm}{\sl P.A.~Lukin}\par}

\vspace{5mm}

P.A. Lukin$^{1}$, V.E. Lyubovitskij$^{2}$

\vspace{5mm}

\noindent
$^1$ Budker Institute of Nuclear Physics and Novosibirsk State University  \\
$^2$ Tuebingen University, Tomsk State University, Tomsk Polytechnic University \\
\vspace{5mm}

Electron-positron collider VEPP-2000~\cite{vepp} has been operating in Budker Institute of Nuclear Physics since 2010. 
Center-of-mass energy ($E_{c.m.}$) range covered by the collider is from threshold of hadron production and up to 2 GeV. 
Special optics, so called ``round beams'', used in the collider construction, allowed to obtain luminosity 
$2\times10^{31}$ cm$^{-2}\cdot$s$^{-1}$ at $E_{c.m.} = $1.8 GeV.   

The general purpose detector CMD-3 has been described in 
detail elsewhere~\cite{sndcmd3}. Its tracking system consists of a 
cylindrical drift chamber (DC)~\cite{dc} and double-layer multiwire 
proportional 
Z-chamber, both also used for a trigger, and both inside a thin 
(0.2~X$_0$) superconducting solenoid with a field of 1.3~T.
The liquid xenon (LXe) barrel calorimeter with 5.4~X$_0$ thickness has
fine electrode structure, providing 1-2 mm spatial resolution~\cite{lxe}, and
shares the cryostat vacuum volume with the superconducting solenoid.     
The barrel CsI crystal calorimeter with thickness 
of 8.1~X$_0$ is placed
outside  the LXe calorimeter,  and the end-cap BGO calorimeter with a 
thickness of 13.4~X$_0$ is placed inside the solenoid~\cite{cal}.
The luminosity is measured using events of Bhabha scattering 
at large angles~\cite{lum}. 

Physics program of the CMD-3 experiment includes the study of the multi-hadron production.
The cross section measurement of the $e^+e^-\to 3(\pi^+\pi^-)$ process in $E_{c.m.} = 1.5~--~2.0$ GeV has been already 
published~\cite{6pi}.  Preliminary results the study 2$(\pi^+\pi^-\pi^0)$ final state has been reported~\cite{4pi2pi0}.

The study of intermediate states which lead to 2$(\pi^+\pi^-\pi^0)$ final state is essential to correctly describe the 
angular correlations between the particles and determine  the registration efficiency of the process under study. 
As it was reported at~\cite{MCWG_Apr14} the intermediate states $\omega(782)3\pi$, $\omega(782)\eta(545)$ and 
$\rho(770)(4\pi)_{S-wave}$ allow satisfactorily describe mass and angular distributions of  the 2$(\pi^+\pi^-\pi^0)$ 
production in $E_{c.m.} = 1.5~--~1.7$ GeV. 

For higher $E_{c.m.}$ it was found possible to describe $\eta(545)$ signal, seen in three-pion mass distribution of 
the 2$(\pi^+\pi^-\pi^0)$, by the process $e^+e^-\to a_0(980)\rho(770)$  with dominant decay of $a_0(980)$ into $\eta(545)\pi$ as it
is described in~\cite{MCWG_Nov14}. The primary Monte-Carlo generator
for the process has been created out and implemented into the CMD-3 experiment Monte-Carlo simulation package. Using
the generator the signal of the $e^+e^-\to a_0(980)\rho(770)$ process has been observed in the experimental data for
2$(\pi^+\pi^-\pi^0)$ final state. 

In the present study, it was made an attempt to describe 2$\pi$- 3$\pi$ and 4$\pi$ mass distributions as well as angular 
correlations for 2$(\pi^+\pi^-\pi^0)$ final state at $E_{c.m.} > 1.8$ GeV by contributions of $\omega(782)3\pi$, 
$a_0(980)\rho(770)$ and $f_0(1370)\rho(770)$ intermediate states. 
Two decay channels of $f_0(1370)$ have been studied $f_0\to\rho^+(770)\rho^-(770)$ and $f_0(1370)\to\pi^+\pi^-2\pi^0$. 
It was found that process $e^+e^-\to f_0(1370)\rho^0(770)$ with subsequent decay $f_0\to\rho^+(770)\rho^-(770)$ reasonably
describe both masses and angular distributions of the $e^+e^-\to 2(\pi^+\pi^-\pi^0)$ process at $E_{c.m.} = 2.0$ GeV. And we
could not described masses and angular distributions of the $e^+e^-\to 2(\pi^+\pi^-\pi^0)$ process at $E_{c.m.} = 1.8$ GeV by
neither $f_0(1370)\to\rho^+(770)\rho^-(770)$  nor by $f_0(1370)\to\pi^+\pi^-2\pi^0$ decays.

\newpage

\subsection{Event Generators at Belle\,II}
\addtocontents{toc}{\hspace{2cm}{\sl T.~Ferber}\par}

\vspace{5mm}

T.~Ferber

\vspace{5mm}
\noindent
DESY,
D-22607 Hamburg, Germany \\
\vspace{5mm}

The next generation B--factory Belle\,II at the upgraded KEKB accelerator, SuperKEKB, is aiming to start data taking in 2017. The broad physics program covers e.g. physics with B and D mesons, $\mu$ and $\tau$ leptons as well as measurements using the method of radiative returns and direct searches for new physics. The expected dataset will exceed the one collected by the predecessor Belle by a factor of 50 and imposes high precision requirements on the used event generators and requires a flexible and powerful software framework. The bulk of the data will be collected at the $\Upsilon(4S)$ resonance, but it is planned to collect sizable data sets also off--resonance and at energies around the narrow resonances $\Upsilon(1S)$, $\Upsilon(2S)$ and $\Upsilon(3S)$.\\

Belle\,II uses a single software framework, basf2 \cite{belle2,belle2b}, for all data processing tasks which runs on standard Linux systems. It is based on a user--defined chain of individual modules where each subsequent module can read data from the preceding modules from a so-called data store. Event generators usually serve as specialized modules that provide four vectors to be fed into the subsequent GEANT4 based detector simulation as well as precision cross section calculations used for normalization. If available from the generator, all mother-daughter relations of both unstable particles and radiative photons are stored and used in Monte Carlo truth matching during analysis. FORTRAN--based generators are interfaced using extern ``C'' functions where the user inputs and generator outputs are provided as FORTRAN common blocks and global C/C++ extern structs of the same name. All generators use a random generator provided by the basf2 framework. The original input interfaces are replaced by Python steering scripts that provide access to the generator options in a uniform and user--friendly way.\\

The available physics generators include BABAYAGA.NLO\,\cite{babayaga}, BHWIDE\,\cite{bhwide}, KKMC4.19\,\cite{kkmc1, kkmc2}, PHOKHARA9.1b\,\cite{phokhara}, KORALW1.51\,\cite{koralw}, AAFH\,\cite{aafh}, BBREM\footnote{C++ implementation based on the original FORTRAN code \cite{bbrem}.} and EvtGen\,\cite{evtgen}. \\
TAUOLA\,\cite{tauola} and PHOTOS\,\cite{photos} are used by EvtGen and KKMC to handle $\tau$ decays and radiative corrections in decays. MadEvent \cite{madevent} is used to simulate New Physics processes. Light quark continuum is modeled using KKMC (hard interaction), PYTHIA8 (fragmentation)\,\cite{pythia8} and EvtGen (decays). Standardized HepEvt or Les Houches event (LHE) format can be read by dedicated input modules. The data from the data store can be obtained at any stage of the module--chain in ROOT or HepEvt file format.\\

In conclusion, all basic event generators are available in basf2 and ready for physics and trigger studies. Future projects will focus on the precision validation and improvements needed to match the demanding precision requirements for luminosity measurements and low multiplicity physics especially at the narrow resonances. A semi--automatic framework to check EvtGen models is under development. Work has started to use the Belle datasets taken off--resonance and at the $\Upsilon(1S)$ to tune PYTHIA8 within the basf2 framework.

\newpage

\subsection{Heavy quark masses form QCD Sum Rules}
\addtocontents{toc}{\hspace{2cm}{\sl P.~Masjuan}\par}

\vspace{5mm}
P.~Masjuan

\vspace{5mm}
\noindent
PRISMA Cluster of Excellence, Institut f\"ur Kernphysik and Helmholtz~Institut~Mainz, Johannes Gutenberg-Universit\"at Mainz,
D-55099 Mainz, Germany \\
\vspace{5mm}

Heavy quark masses are interesting not only for being fundamental parameters not predictable within the Standard Model (SM), but for their implications on many phenomenological scenarios as well. Being a free parameter in the SM, their values should be extracted from experimental information which already demands an important synergy between both experiment and theory. Beyond that, the heavy quark mass enters into decays, from Higgs (where the mass enters squared) to $B$ decays (where enters with the 5th exponential). For precision physics, an accurate knowledge of that important input is mandatory, specially if one foresees Higgs branching ratios at the per mil level.

From the different methods to extract the quark mass from experimental data, we follow in Ref.~\cite{EMS} the Sum Rule's one~\cite{Novikov:1977dq,Shifman:1978bx,Shifman:1978by,Kuhn:2007vp}. The method relates, thanks to the optical theorem, the transverse part of the correlator of two heavy-quark vector currents denoted by $\hat\Pi_q(t)$ ---where the caret indicates $\overline{\rm MS}$ subtraction--- with integrals over the measured ratio $R$, the ratio between the $\sigma(e^+e^- \to hadrons)$ over the $\sigma(e^+e^- \to \mu^+ \mu^-)$. 

The $\hat\Pi_q(t)$ can be calculated order by order within perturbative QCD as an expansion of the strong coupling $\alpha(s)$, and obeys the subtracted dispersion relation \cite{Erler:2002bu} 
\begin{equation}
  12 \pi^2 \frac{\hat\Pi_q (0) - \hat\Pi_q (-t)}{t} 
  =
  \int_{4 m_q^2}^\infty {{\rm d} s\over s} {R_q(s)\over s + t}
  \, ,
\label{SR}
\end{equation}
where $R_q(s) = 12 \pi \mbox{Im} \hat\Pi_q(s)$, and $m_q$ is the mass of the heavy quark that we want to determine. In the limit $t\rightarrow 0$, Eq.~(\ref{SR}) coincides with the first moment, ${\cal M}_1$, of $\hat\Pi_q(t)$. In general, there is a sum rule for each higher moment as well~\cite{Novikov:1977dq,Shifman:1978bx,Shifman:1978by,Kuhn:2007vp}: 
\begin{equation}
  {\cal M}_n 
  := 
  \left.{12\pi^2\over n !} {d^n\over d t^n} 
  \hat\Pi_q(t) \right|_{t=0} 
  = 
  \int_{4 m_q^2}^\infty {{\rm d} s\over s^{n+1}} R_q(s) \, .
\label{SRder}
\end{equation}
Taking the opposite limit, i.e.\ $t \rightarrow \infty$, of Eq.~(\ref{SR}) after multiplying with $t$ and with a proper regularization, one can define a sum rule for the $0^{th}$ moment \cite{Erler:2002bu}. At a given, fixed order of pQCD, the required regularization can be obtained by subtracting the zero-mass limit of $R_{q}$, which we denote $3 Q_q^2 \lambda_1^q(s)$ with $Q_q$ the quark charge. 
$\lambda_1^q(s)$ is known up to ${\cal O}(\alpha_s^3)$ \cite{Chetyrkin:2000zk}.

By the optical theorem, $R_{q}$ can be related to the measurable cross section for heavy-quark production in $e^+e^-$ annihilation. We assume that below the threshold, the cross section is determined by a small number of narrow Breit-Wigner resonances \cite{Novikov:1977dq}
\begin{equation}
  R_q^{\rm Res}(s) 
  = 
  \frac{9\pi}{\alpha_{\rm em}^2(M_R)} 
  M_R \Gamma_R^e\delta(s-M_R^2)\, .
\label{Rres}
\end{equation}
The masses $M_R$ and electronic widths $\Gamma^e_R$ of the resonances are collected from the PDG~\cite{Agashe:2014kda} 
and $\alpha_{\rm em}(M_R)$ is the running fine structure constant at the resonance. Then, $R_q(s) = R_q^{\rm Res}(s) +R_q^{\rm cont}(s)$, with $R_q^{\rm cont}(s)$ accounting for the continuum production region.

Invoking global quark-hadron duality, we also assume that continuum production can be described on average by the simple ansatz \cite{Erler:2002bu} (which coincides with the reconstruction of the correlator at the given order~\cite{Kiyo:2009gb,Hoang:2008qy,Greynat:2011zp})
\begin{equation} 
  R_q^{\rm cont}(s)
  = 3 Q^2_q \lambda^q_1 (s) 
  \sqrt{1 - {4\, \hat{m}_q^2 (2 M) \over s^\prime}} 
  \left[ 1 + \lambda^q_3 {2\, \hat{m}_q^2(2 M) \over s^\prime} 
  \right]
\label{ansatz}
\end{equation}
where $s' = s + 4(\hat{m}_q^2(2M) - M^2)$ and $\hat{m}_q$ is the running $\overline{\rm MS}$ heavy-quark mass, evaluated at ${\cal O}(\alpha_s^3)$~\cite{Vermaseren:1997fq} at the scale $2M$, $M$ taken as mass of the lightest pseudoscalar heavy meson. $\lambda^q_3$ is a constant to be determined. $R_q^{\rm cont}(s)$ interpolates smoothly between the threshold and the onset of open heavy-quark pair production. It coincides asymptotically with the prediction of pQCD for massive quarks. 
Two parameters are free to be determine after fits to the experimental data by BES~\cite{Bai:1999pk, 
Bai:2001ct,Ablikim:2006mb,Ablikim:2006aj,Ablikim:2009ad}, 
and CLEO~\cite{CroninHennessy:2008yi} on the open charm region: $\lambda_3$ and $\hat{m}_q$, for which two different sum rules will be used, the $0th$ and the $2nd$. In Ref.~\cite{EMS}, not only the central values of the heavy masses are given, but also an exhaustive scrutiny of the many error sources is also presented.

In this talk, we presented our preliminary results on the study of the charm quark mass value within the framework of QCD Sum Rules at ${\cal O} (\alpha_s^3)$. Our final numbers are still preliminary and are not reported in the present discussion.
We profit from the discussion on our results with our experimental colleagues present in this workshop. We are thankful to the organizers for providing such an excellent environment for that kind of interactions.

\newpage

\subsection{Current status of {\tt carlomat\_3.0}, an automatic tool for low 
       energetic electron-positron annihilation into hadrons}
\addtocontents{toc}{\hspace{2cm}{\sl K.~Kolodziej}\par}

\vspace{5mm}

K.~Ko\l odziej

\vspace{5mm}

\noindent
Institute of Physics, University of Silesia, ul. Uniwersytecka 4, PL-40\,007
Katowice, Poland\\
\vspace{5mm}

In the energy range below  the $J/\psi$ threshold, processes of 
$\epm$-annihilation to hadrons cannot be described in the framework of
perturbative QCD. The scalar electrodynamics (sQED) which was implemented 
in {\tt carlomat\_2.0} \cite{carlomat2} does not describe them in a 
satisfactory manner either. 
The theoretical frameworks usually used for the description of such processes 
are the Resonance Chiral Theory (R$\chi$T) or Hidden Local Symmetry (HLS) model 
which were proven to be equivalent in this context \cite{equiv}.
Both the R$\chi$T and HLS model involve, among others, the photon--vector 
meson mixing and a substantial number of vertices of a rather complicated 
Lorentz tensor structure that is present neither in the Standard Model (SM)
nor sQED. Even at low energies, the final state of $\epm \to {\rm hadrons}$
may consist of several particles, 
such as pions, kaons, or nucleons with one or more photons, or light 
fermion pairs such as $e^+e^-$, or $\mu^+\mu^-$ receive contribution
of a big number of the Feynman diagrams. A new version of a program
{\tt carlomat} \cite{carlomat}, tagged with index 3.0 \cite{carlomat3}, is just 
dedicated to the description of such multiparticle processes involving
hadrons in a fully automatic way.

Implementation of the photon--vector meson mixing required substantial 
modification of the code generation part of {\tt carlomat}. Further
changes in that part of the program were needed in order to incorporate 
calls to new subroutines for computation of the helicity amplitudes of
the building blocks and complete Feynman diagrams which contain new
interaction vertices and mixing terms. 
Moreover, a number of new subroutines  for computation of the helicity 
amplitudes involving the Feynman interaction vertices of the R$\chi$T or
HLS model \cite{Benayoun} with the Lorentz tensor structures, which are 
not present either in the SM or in the effective models
implemented in the former version of the program, were written and
many subroutines have been modified in order to incorporate the $q^2$-dependent
couplings and vector meson widths. 

In order to give the user a better control of the implemented effective models,
a number of new options have been introduced in the Monte Carlo computation
part of the program. They include a possibility of choosing different formulae
for the $q^2$-dependent couplings, where the four momentum transfer $q$ is 
determined automatically at the 
stage of code generation from the four momentum conservation in the 
corresponding interaction vertex, or for the $s$-dependent widths. 
It is also possible to modify weights and phases of each of
the photon--vector meson mixing terms implemented in the program, which 
should help to find out
the dominant production mechanisms of different hadronic channels. Another
important option offers a possibility to
test the electromagnetic gauge invariance for processes involving one ore more
photons.

{\tt carlomat\_3.0} can be downloaded from CPC Program Library or from the web
page: {\tt http://kk.us.edu.pl/carlomat.html.}

{\bf Acknowledgement:} This project was supported in part with financial 
resources of the Polish National Science Centre (NCN) under grant decision 
No. DEC-2011/03/B/ST6/01615. The author is grateful to Fred Jegerlehner for
providing the Feynman rules of the HLS model.

\newpage

\subsection{A first glance towards a dispersive formalism for $\gamma^{*} \gamma \rightarrow \pi \pi$}
\addtocontents{toc}{\hspace{2cm}{\sl P.~Masjuan}\par}

\vspace{5mm}
P.~Masjuan

\vspace{5mm}
\noindent
PRISMA Cluster of Excellence, Institut f\"ur Kernphysik and Helmholtz~Institut~Mainz, Johannes Gutenberg-Universit\"at Mainz,
D-55099 Mainz, Germany \\
\vspace{5mm}

We present a first step towards a dispersive formalism for the $\gamma^{*} \gamma \rightarrow \pi \pi$ process. This process is not only interesting by its own, as it will be measured for both one- and two-virtual photons in BESIII and Belle with high precision and encodes at once several interesting theoretical aspects (gauge invariance, final-state interactions, form factors), but also for its potential relation to the hadronic light-by-light scattering (see~\cite{Masjuan:2014rea,Adlarson:2014hka,Benayoun:2014tra} and references therein) and the generalized pion polarizabilities. 

Our goal is to provide a friendly useful MonteCarlo parameterization based on dispersion relations~\cite{Pennington,Drechsel:1999rf,GarciaMartin:2010cw,Hoferichter:2011wk,Dai:2014zta}  while keeping its essential ingredients~\cite{MSPV}. This demands identifying the crucial pieces of information that allow for a reliable, albeit not complete, description of the current~\cite{belle_pic,belle_pin} and forthcoming data at BESIII. The analysis of the recent works~\cite{GarciaMartin:2010cw,Hoferichter:2011wk,Dai:2014zta}, and taking into account only the $\pi\pi$ channel (neglecting then any inelasticity up to almost $1$ GeV), suggests to neglect the $K\bar{K}$ channel, while keeping left-hand cut contributions beyond the one-pion exchange in a manageable way~\cite{Moussallam:2013una}. 

With this ingredients at hand, we proceed to describe an unsubtracted dispersion relation (DR) as in Ref.~\cite{Drechsel:1999rf} but with both S- and D-waves unitarized using, thanks to the Fermi-Watson theorem, the corresponding $\pi\pi$ phase-shifts, solutions of which are taken from Ref.~\cite{GarciaMartin:2011cn} for both isospin $I=0,2$  (the only accessible with two incoming photons). Since, however, we want to consider the elastic $\pi\pi$ channel only, we should modify the phase shift around the $KK$ threshold for emulating, without employing a coupled channel formalism, the amplitude's phase. The coupled channel is known~\cite{GarciaMartin:2010cw,Dai:2014zta} but difficult to implement with virtual photons~\cite{Moussallam:2013una}, and hard to simplify for a user friendly MonteCarlo generator. For that we use the proposal of Ref.~\cite{Moussallam:2013una} to define a piecewise function. The high energy behavior of the phase-shift, beyond the parameterization provided in~\cite{GarciaMartin:2011cn}, is matched to the Regge behavior~\cite{Dai:2014zta} using the latest Regge analysis of the PDG data~\cite{Masjuan:2012gc}.

Virtual photons imply extending the DR for $\gamma \gamma \rightarrow \pi \pi$, which has two independent helicity amplitudes, to three (one-virtual photon) and five (two-virtual photons) independent amplitudes. For the one-virtual photon discussed in this talk, the new amplitude yields an enhancement on the cross section around the threshold of the same order as the transversal amplitudes. Such effect helps to compensate the decrease of the cross section due to the flux factor suppression in presence of photon virtualities together with the $\gamma-\pi^0$ form factor (and the soft-photon limits~\cite{Moussallam:2013una}).

As we just said, as an ingredient we need the coupling of the virtual photon with the pion (given by the vector form factor). We use a data driven parameterization specially suited for low energies from Ref.~\cite{Masjuan:2008fv} . Around the $f_2(1275)$ tensor resonance, which shows up in the the $\gamma \gamma \rightarrow \pi \pi$ process around $1.3$GeV, the amplitude can be worked out with the help of a Breit-Wigner model for the $f_2$~\cite{Drechsel:1999rf}, or using the appropriate $\pi\pi$ phase-shifts together with the left-hand-cuts of the amplitudes (i.e, vector and axial-vector contributions in the $t-$channel exchange). Even though we are concerned with the low-energy sector of the process, we also include the effects of the $f_2$ in our formalism since they spread all over the cross section. Within the former approach, the $f_2$ coupling to the virtual photon can be parameterized in terms of vector meson dominance ideas~\cite{Masjuan:2012sk}, using the $\eta$ and $\eta'$ form factors~\cite{Escribano:2013kba}. For the later, the coupling of the virtual vector with the pion and the (left-hand) vector/axial-vector, in terms of meson dominance~\cite{Masjuan:2012sk,Moussallam:2013una}.
For a photon virtuality around $Q^2=0.5$ GeV$^2$, the typical intensity of the integrated cross section is of around 30 $nb$~\cite{MSPV}, measurable at BESIII.

\newpage

\subsection{ Current status of two and three pion 
decay modes within TAUOLA}
\addtocontents{toc}{\hspace{2cm}{\sl O.~Shekhovtsova }\par}

\vspace{5mm}

O.~Shekhovtsova 

\vspace{5mm} 
\noindent
Institute of Nuclear Physics PAN ul. Radzikowskiego 152 31-342 Krakow, Poland \\ 
Kharkov Institute of Physics and Technology   61108, Akademicheskaya,1, Kharkov, Ukraine \\ 
\vspace{5mm} 

Since 90' years Tauola is the main Monte Carlo generator to simulate tau-lepton decays \cite{Jadach:1993hs}. It has been used by the collaborations Aleph \cite{Buskulic:1995ty}, Cleo  \cite{Asner:1999kj}, at both B-factories (BaBar \cite{Nugent:2013ij} and Belle \cite{Fujikawa:2008ma}) as well at LHC \cite{Aad:2012tfa} experiments for tau decay data analysis. 
In view of the forthcoming Belle-II project \cite{Abe:2010gxa} it is important to revise the Tauola context in details.
In this notes we discuss the status of the predominant hadronic tau-lepton decay modes: two pion ($Br \simeq 25.52\%$) and three pion ($Br \simeq 18.67\%$) decay channels. 

In the general case, a hadronic current of two-meson tau-lepton decay mode depends on both vector and scalar form factors. However, in the isospin  symmetry limit, $m_{\pi^-} = m_{\pi^0}$, the scalar form factor vanishes for two-pion decay mode and the current is described by the vector form-factor only. Currently Tauola has four parameterizations for the vector form-factor: 
\begin{itemize} 
\item Kuhn-Santamaria (KS) parametrization \cite{Kuhn:1990ad}; 
\item Gounaris-Sakurai (GS) parametrization used by Belle \cite{Fujikawa:2008ma}, Aleph and Cleo collaborations; 
\item parametrization based on the Resonance Chiral Lagrangian (RChL) \cite{SanzCillero:2002bs}; 
\item combined parametrization (combRChL) that applies dispersive approximation at low energy and modified RChL result at high energy \cite{Dumm:2013zh}. 
\end{itemize}  
In all cases, except for the RChL parametrization, the pion form factor is given by interfering amplitudes from the known isovector meson resonances $\rho(770)$, $\rho'(1450)$ and $\rho''(1700)$ with relative strengths $1$, $\beta$ and $\gamma$.  Although one could expect from the quark model that $\beta$ and $\gamma$ are real, their phases   are left free in the fits. 
In the case of the RChL parametrization only the $\rho(700)$ and $\rho'(1450)$ contributions are included, with the relative $\rho'$ strength being a real parameter. 
For the energy-dependence of $\rho(770)$ two-pion and two kaon loop contributions are included for RChL and combRChl parameterizations, whereas in the case of KS and GS the $\rho$ width is approximated by the two pion loops only. The $\rho'$ and $\rho''$ widths include two-pion loops only for all parameterizations.  

Results of the fit to the Belle data \cite{Fujikawa:2008ma} are presented in Fig.~1. The best fit is within the GS pion form-factor and the worst one is within the RChL one. To check the influence of the $\rho''(1700)$ on the RChL result, the $\rho''$ resonance has been included in the same way as it was done for $\rho'$; however, this inclusion has not improved the result. Therefore, we conclude that missing loop contributions could be responsible for the disagreement and that the phases of $\beta$ and $\gamma$ 
might mimic the missing loop contribution. This idea will be checked by adding, first, a four-pion loop contributions. 

\begin{figure}[h]
\includegraphics[scale = .27]{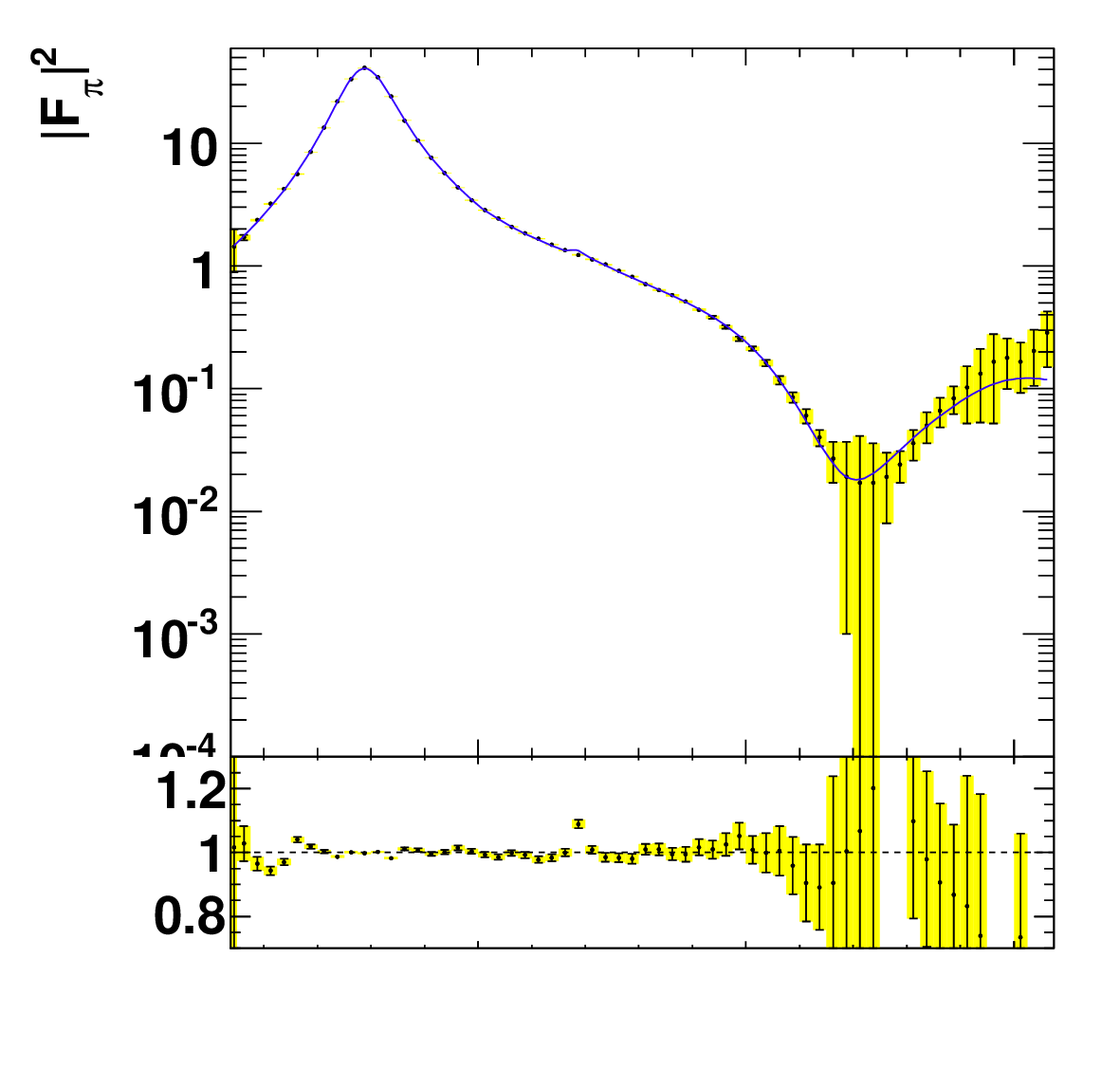}
\includegraphics[scale = .27]{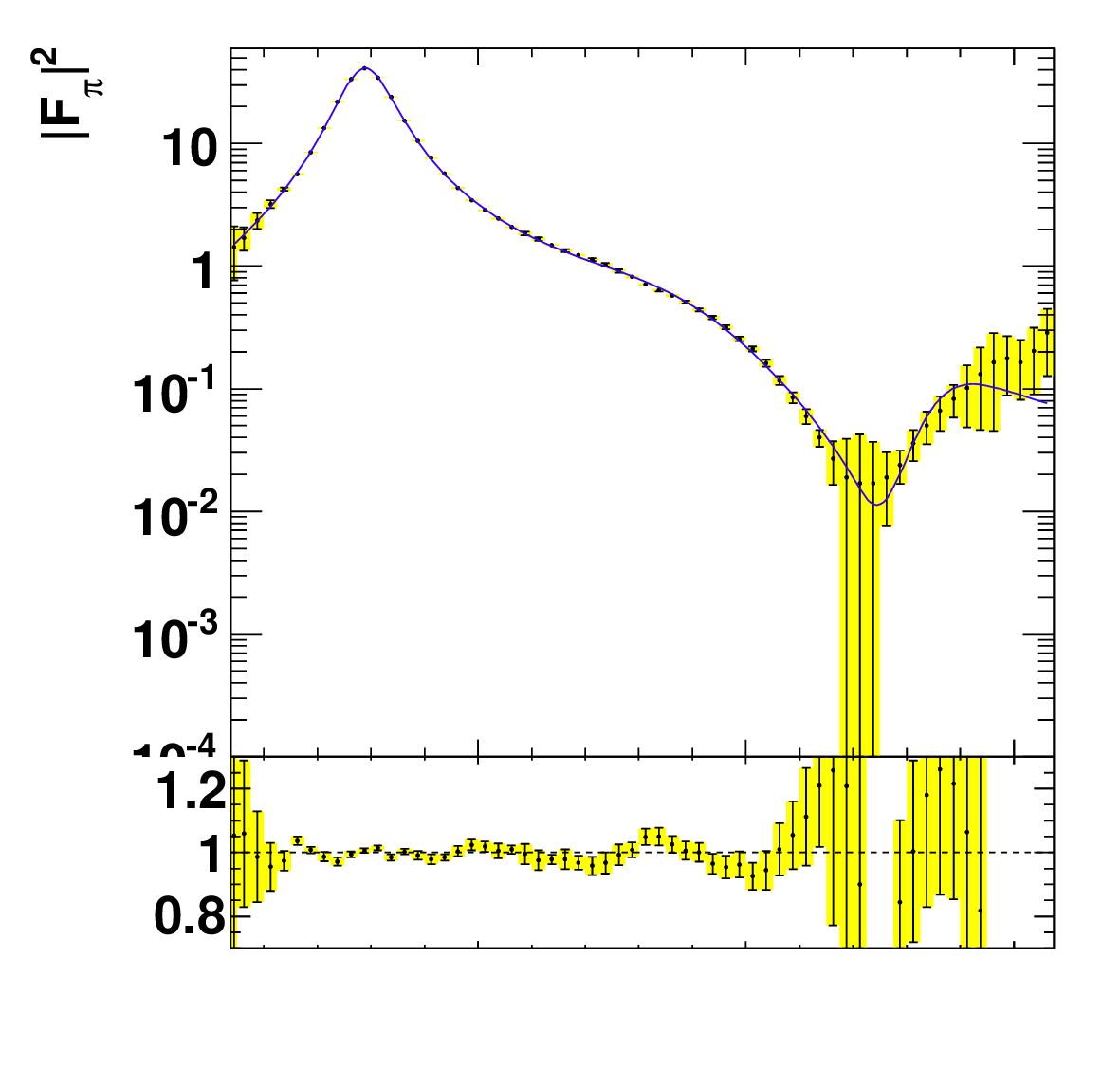}
\includegraphics[scale = .27]{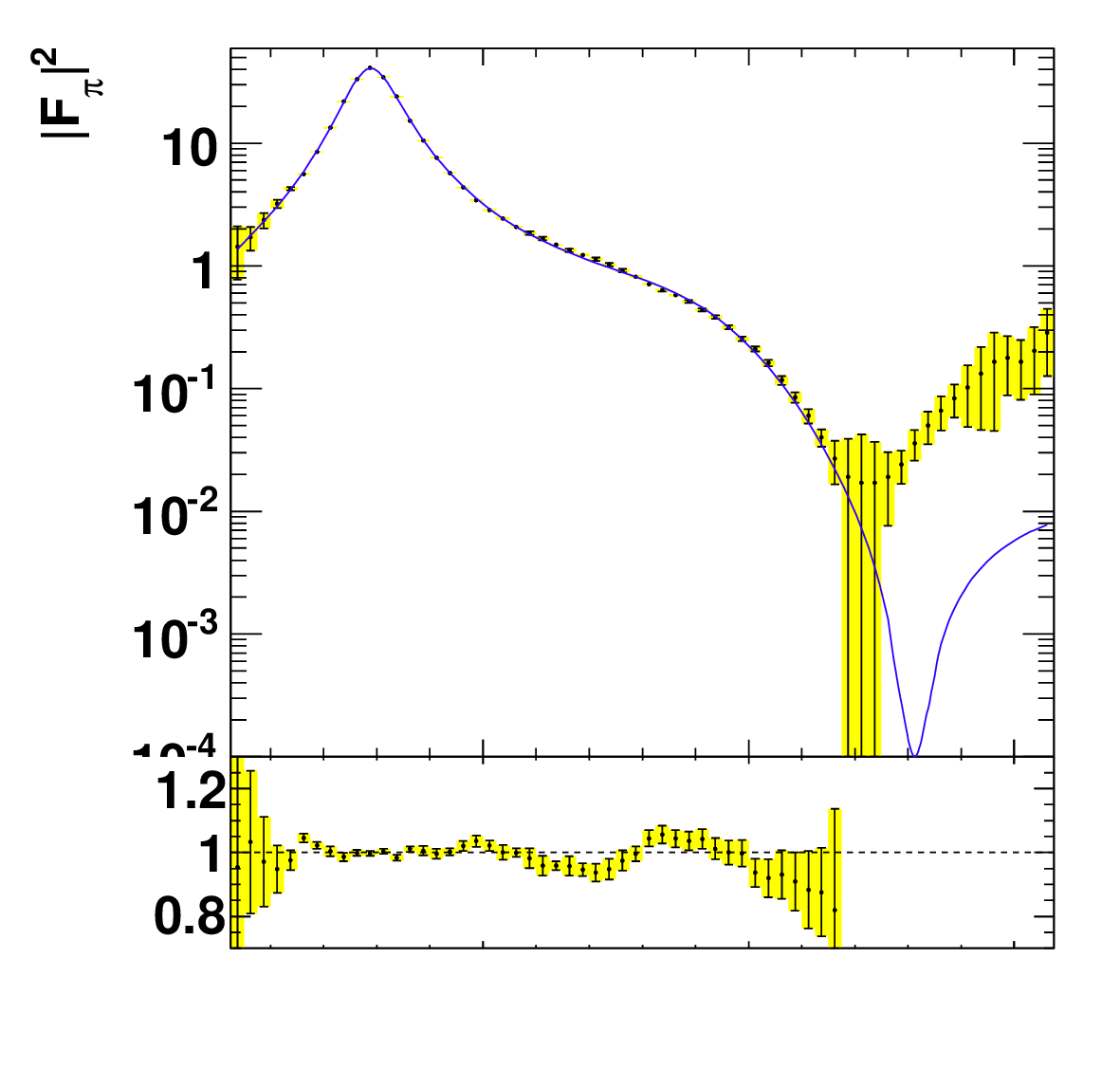}
\caption{The pion form factor fit to Belle data \cite{Fujikawa:2008ma}: the GS parametrization (left panel), the RChL parametrization (at centre), right panel is with the combRChL parametrization (right panel).}
\end{figure}
When the $\tau$ lepton decays into three hadrons and a neutrino, the predominant decay mode involves three pions. In the general case Lorentz invariance determines the decomposition of the hadronic current for a three hadron final state in  terms of  five Lorentz invariant structures \cite{Shekhovtsova:2012ra} multiplied by hadronic form factors. Among the three hadronic form factors which correspond 
to the axial-vector part of the hadronic tensor, only two are independent. In Tauola the following three pion form factors are available 
\begin{itemize} 
\item CPC version \cite{Jadach:1993hs}, that includes only the dominant $a_1 \to \rho\pi$ mechanism production  
\item Cleo parametrization with equal currents for $\pi^0\pi^0\pi^-$ and $\pi^-\pi^-\pi^+$ modes based on \cite{Asner:1999kj} 
\item Cleo parametrization with the $\pi^0\pi^0\pi^-$ current \cite{Asner:1999kj} and the $\pi^-\pi^-\pi^+$ current from  the unpublished Cleo analysis \cite{Shibata:2002uv} (the currents coincide for vector and axial-vector intermediate states and are not equal when the scalar and tensor resonances are included) 
\item modified RChL parametrization \cite{Nugent:2013hxa} 
\end{itemize} 

The Cleo parametrization results have not yet been compared with the BaBar preliminary data \cite{Nugent:2013ij} and it will be a task of future work. Whereas two- and three-pion spectra calculated on the base of the modified RChL parametrization have been fitted to the BaBar preliminary results. Discrepancy between  theoretical spectra and experimental data can be explained by    missing resonances in the model, such as the axial-vector resonance $a_1'(1600)$, the scalar resonance $f_0(980)$ and the tensor resonance $f_2(1270)$. Inclusion of these resonances in the RChL framework will be a future task. 

Comparison of the $\pi^-\pi^-\pi^+$ current in the framework of the modified RChL with the ChPT result has demonstrated that the scalar resonance contribution has to be corrected to reproduce the low energy ChPT limit. The corresponding calculation is under work. 
The same type of comparison has to be done for both version of the Cleo $\pi^-\pi^-\pi^+$ currents described above. 
 
This research was supported in part from funds of Foundation of Polish Science grant POMOST/2013-7/12, that is co-financed from European Union, Regional 
Development 
Fund, and by Polish National Science 
Centre under decisions  DEC-2011/03/B/ST2/00107.

\newpage

\subsection{A new approach to evaluate the leading hadronic corrections to the muon \boldmath $g$-2\unboldmath}
\addtocontents{toc}{\hspace{2cm}{\sl  G.~Venanzoni}\par}

\vspace{5mm} 
C.\ M.\ Carloni Calame$^1$, M. Passera$^2$, L.\ Trentadue$^3$,\underline{G. Venanzoni}$^{4}$\\
$^1$ Dipartimento di Fisica, Universit\`a di Pavia, Pavia, Italy\\
$^2$ INFN, Sezione di Padova, Padova, Italy \\
$^3$ Dipartimento di Fisica e Scienze della Terra 
``M.\ Melloni''\\Universit\`a di Parma, Parma, Italy and \\ INFN, Sezione di Milano Bicocca, Milano, Italy \\
$^4$ Laboratori Nazionali di Frascati dell'INFN, 00044 Frascati, Italy\\ 
\vspace{5mm}

The long-standing discrepancy between experiment and the Standard Model (SM) prediction of $a_{\mu}$, the muon anomalous magnetic moment, has kept the hadronic corrections under close scrutiny for several years~\cite{Bennett:2006fi,Jegerlehner:2009ry,Reviews,Jegerlehner:2008zza}. In fact, the hadronic uncertainty dominates that of the SM value and is comparable with the experimental one. When the new results from the $g$-2 experiments at Fermilab and J-PARC will reach the unprecedented precision of 0.14 parts per million (or better)~\cite{Grange:2015fou,Venanzoni:2014ixa,Saito:2012zz}, the uncertainty of the hadronic corrections will become the main limitation of this formidable test of the SM.

An intense research program is under way to improve the evaluation of the leading order (LO) hadronic contribution to $a_{\mu}$, due to the hadronic vacuum polarization correction to the one-loop diagram~\cite{Venanzoni:2014wva,Fedotovich:2008zz}, as well as the next-to-leading order (NLO) hadronic one. The latter is further divided into the $O(\alpha^3)$ contribution of diagrams containing hadronic vacuum polarization insertions~\cite{Krause:1996rf}, and the leading hadronic light-by-light term, also of $O(\alpha^3)$~\cite{Jegerlehner:2009ry,HLBL,HLBLfuture}. Very recently, even the next-to-next-to leading order (NNLO) hadronic contributions have been studied: insertions of hadronic vacuum polarizations were computed in~\cite{Kurz:2014wya}, while hadronic light-by-light corrections have been estimated in~\cite{Colangelo:2014qya}.

The evaluation of the hadronic LO contribution $a_{\mu}^{\scriptscriptstyle \rm HLO}$ involves long-distance QCD for which perturbation theory cannot be employed. However, using analyticity and unitarity, it was shown long ago that this term can be computed via a dispersion integral using the cross section for low-energy hadronic $e^+ e^-$ annihilation~\cite{BM61GDR69}. At low energy this cross-section is highly fluctuating due to resonances and particle production threshold effects.

An alternative determination of $a_{\mu}^{\scriptscriptstyle \rm HLO}$ can be obtained measuring the effective electromagnetic coupling in the space-like region extracted from Bhabha ($e^+ e^- \to e^+ e^-$) 
scattering data~\cite{Calame:2015fva}. 
As vacuum polarization in the space-like region is a smooth function of the squared momentum transfer, the accuracy of its determination is only limited by the statistics and by the control of the systematics of the experiment. Also, as at flavor factories the Bhabha cross section is strongly enhanced in the forward region, we will argue that a space-like determination of $a_{\mu}^{\scriptscriptstyle \rm HLO}$ may not be limited by statistics and, although challenging, may become competitive with standard results obtained with the dispersive approach via time-like data.

\newpage

\section{List of participants}

\begin{flushleft}
\begin{itemize}
\item Carlo M. Carloni Calame, University of Pavia, {\tt carlo.carloni.calame@pv.infn.it}
\item Francesca Curciarello, University of Messina, {\tt fcurciarello@unime.it}
\item Henryk Czy\.z, University of Silesia, {\tt henryk.czyz@us.edu.pl}
\item Veronica de Leo, INFN Sezione de Roma 3, {\tt veronica.deleo@roma3.infn.it}
\item Simon Eidelman, BINP, Novosibirsk State University, {\tt eidelman@mail.cern.ch}
\item Torben Ferber, DESY, {\tt torben.ferber@desy.de}
\item Andreas Hafner, {\tt hafner@slac.stanford.edu}
\item Wolfgang Kluge, Karlsruhe Institue of Technology, {\tt wolfgang.kluge@partner.kit.edu}
\item Karol Ko\l odziej, University of Silesia, {\tt karol.kolodziej@us.edu.pl}
\item Andrzej Kupsc, University of Uppsala, {\tt Andrzej.Kupsc@physics.uu.se}
\item Peter Lukin, BINP and NSU, {\tt P.A.Lukin@inp.nsk.su}
\item Giuseppe Mandaglio, University of Messina, {\tt gmandaglio@unime.it}
\item Pere Masjuan, Universt\"at Mainz, {\tt masjuan@kph.uni-mainz.de}
\item Federico Nguyen, ENEA - Agency for New Technologies, Energy and Sustainable Economic Development, {\tt federico.nguyen@enea.it}
\item Benjamin Oberhof, LNF, {\tt benjamin.oberhof@lnf.infn.it}
\item Chirsoph Florian Redmer, Universt\"at Mainz, {\tt redmer@kph.uni-mainz.de}
\item Olga Shekhovtsova, IFJ PAN, {\tt Olga.Shekhovtsova@lnf.infn.it}
\item Graziano Venanzoni, LNF, {\tt Graziano.Venanzoni@lnf.infn.it}
\item Ping Wang, Institute of High-energy Physics, Beijing, {\tt wangp@ihep.ac.cn}
\end{itemize}
\end{flushleft}


\begin{thebibliography}{99}
\bibitem{Actis:2010gg}
  S.~Actis {\it et al.}  [Working Group on Radiative Corrections and Monte Carlo Generators for Low Energies Collaboration],
  Eur.\ Phys.\ J.\ C {\bf 66} (2010) 585
  [arXiv:0912.0749 [hep-ph]].
\bibitem{Czyz:2013zga}
  P.~Masjuan, G.~Venanzoni, H.~Czy\.z, A.~Denig, M.~Vanderhaeghen, G.~Venanzoni, A.~Denig and S.~Eidelman {\it et al.},
  arXiv:1306.2045 [hep-ph].
 \bibitem{Czyz:2013sga}
  H.~Czy\.z, S.~Eidelman, G.~V.~Fedotovich, A.~Korobov, S.~E.~MŸller, A.~Nyffeler, P.~Roig and O.~Shekhovtsova {\it et al.},
  arXiv:1312.0454 [hep-ph].
\bibitem{vanderBij:2014mxa}
  J.~J.~van der Bij, H.~Czy\.z, S.~Eidelman, G.~Fedotovich, T.~Ferber, V.~Ivanov, A.~Korobov and Z.~Liu {\it et al.},
  arXiv:1406.4639 [hep-ph].
\bibitem{Carloni:2014mpa}
  C.~M.~Carloni {\it et al.},
  arXiv:1412.7714 [hep-ph].
 \end{thebibliography}

\begin{thebibliography}{99}
\vspace{-3mm}

\bibitem{ACHASOV:2014nra}
M.N. Achasov et al. (CMD-3 and SND Collaborations),
Int.J.Mod.Phys.Conf.Ser. {\bf 35}, 1460388 (2014).

\bibitem{Akhmetshin:2014lea}
R.R. Akhmetshin et al. (CMD-3 Collaboration),
JINST {\bf 9}, C09003 (2014).

\bibitem{Aulchenko:2014vkn}
V.M. Aulchenko et al. (SND Collaboration), 
Phys.\ Rev.\  D {\bf 91}, 052013 (2015).

\bibitem{snd3pi15} 
V.M. Aulchenko et al.  (SND Collaboration), 
J.\ Exp.\ Theor.\ Phys.\ {\bf 121}, 34 (2015). 

\bibitem{Lukin:2015vsa}
P.A.Lukin et al. (CMD-3 Collaboration),
Phys.\ Atom.\ Nucl.\  {\bf 78}, 353 (2015).

\bibitem{Akhmetshin:2013xc}
R.R. Akhmetshin et al. (CMD-3 Collaboration),
Phys.\ Lett.\ B {\bf 723}, 82 (2013). 

\bibitem{Kozyrev:2015usa}
E.A. Kozyrev et al. (CMD-3 Collaboration), 
Phys.\ Atom.\ Nucl.\  {\bf 78}, 358 (2015).

\bibitem{Achasov:2013btb}
M.N. Achasov et al. (SND Collaboration),
Phys.\ Rev.\ D {\bf 88}, 054013 (2013).

\bibitem{Achasov:2013eli}
M.N. Achasov  et al. (SND Collaboration),
Phys.\ Rev.\ D {\bf 90}, 032002 (2014).

\bibitem{Achasov:2014ncd}
M.N. Achasov  et al. (SND Collaboration), 
Phys.\ Rev.\  D {\bf 90}, 112007 (2014). 

\bibitem{Antonelli:1998fv}
A. Antonelli et al. (FENICE Collaboration),
Nucl.\ Phys.\ B {\bf 517}, 3 (1998).

\bibitem{AKHMETSHIN:2014qta}
R.R. Akhmetshin et al. (CMD-3 Collaboration),  
Int.J.Mod.Phys.Conf.Ser. {\bf 35}, 1460457 (2014).

\bibitem{cplus}
P. Vorobev et al.  (ND Collaboration), 
Sov.\ J.\ Nucl.\ Phys.\ {\bf 48}, 273 (1988).

\bibitem{Akhmetshin:2014hxv}
R.R. Akhmetshin et al. (CMD-3 Collaboration),
Phys.\ Lett.\ B {\bf 740}, 273 (2015).

\bibitem{Achasov:2015mek}
M.N. Achasov  et al. (SND Collaboration), 
Phys.\ Rev.\  D {\bf 91}, 092010 (2015).

\bibitem{Agakishiev:2013fwl}
G. Agakishiev et al. (HADES Collaboration),
Phys.\ Lett.\ B {\bf 731}, 265 (2014). 

\bibitem{Czyz:2013sga}
H. Czy\.{z} et al.,
arXiv:1312.0454.

\bibitem{vanderBij:2014mxa}
J.J. van der Bij et al.,
arXiv:1406.4639.


\end{thebibliography}

\begin{thebibliography}{99}
\setlength{\itemsep}{0cm}

\bibitem{bnl}
G.W.~Bennett{\em et al.}, Phys. Rev. D {\bf 73}, 072003 (2006). 

\bibitem{davier}
M.~Davier {\em et al.}, Europ. Phys. J. C {\bf 71}, 1515 (2011).

\bibitem{teubner}
K.~Hagiwara  {\em et al.}, J.Phys. G {\bf 38}, 085003 (2011).

\bibitem{fermilab} G.~Venanzoni, (Fermilab E989 Coll.), Nucl.~Phys.~Proc.~Suppl.  {\bf 225-227}, 277-281 (2012).

\bibitem{jparc} T.~Mibe, (J-PARC g-2 Collaboration), Nucl.~Phys.~Proc.~Suppl. {\bf 218}, 242-246 (2011).

\bibitem{brodsky}
S.J. Brodsky and E. de Rafael, Phys. Rev. {\bf 168}, 1620 (1968).

\bibitem{kloe:all}
F.~Ambrosino {\em et al.} (KLOE~Collaboration), Phys. Lett. B 670, 285 (2009),  
Phys.Lett. B700,  102-110 (2011), 
Phys.Lett. B720, 336-343 (2013). 

\bibitem{babar:data2}
B.~Aubert {\em et al.} (\babar~Collaboration), PRD {\bf 70}, 072004 (2004), PRD {\bf 71}, 052001 (2005), 
PRD {\bf 73}, 052003 (2006), 
PRD {\bf 73}, 012005 (2006), 
PRD {\bf 76}, 092006 (2007), 
PRD {\bf 76}, 092005 (2007), 
PRD {\bf 76}, 012008 (2007), 
PRD {\bf 77}, 092002 (2008), 
PRL {\bf 103}, 231801 (2009),
PRD {\bf 85}, 112009 (2012),
PRD {\bf 86}, 032013 (2012),
PRD {\bf 88}, 032013 (2013),
PRD {\bf 89}, 092002 (2014).

\bibitem{psiprprp}
Chinese Physics CPC 37, 123001 (2013).
\bibitem{baba35}
Babayaga.3.5: Information at http://www.lnf.infn.it/wg/sighad/.
\bibitem{babanlo}
Babayaga@NLO: G. Balossini, C. M. Carloni Calame, G. Montagna, O. Nicrosini and F. Piccinini, Nucl. Phys. B 758, 227 (2006).
\end{thebibliography}

\begin{thebibliography}{99}
\setlength{\itemsep}{0cm}

\bibitem{JegNyf} F.~Jegerlehner and A.~Nyffeler, Phys. Rept. {\textbf 477}, 1-110 (2009).

\bibitem{davier} M.~Davier {\em et al.}, Eur. Phys. J. {\textbf C 71}, 1515 (2011).

\bibitem{kloe:data} F.~Ambrosino {\em et al.}, [KLOE~Collaboration], Phys. Lett. {\textbf B 670}, 285 (2009).\\  
                    F.~Ambrosino {\em et al.}, [KLOE~Collaboration], Phys. Lett. {\textbf B 700},  102-110 (2011).\\ 
                    D.~Babusci {\em et al.}, [KLOE~Collaboration], Phys. Lett. {\textbf B 720}, 336-343 (2013). 

\bibitem{babar:data} B.~Aubert {\em et al.}, [BaBar~Collaboration], Phys. Rev. Lett. \textbf{103}, 231801 (2009).
                     J.~P.~Lees {\em et al.}, [BaBar Collaboration], Phys. Rev. \textbf{D 86}, 032013 (2012)
\bibitem{BES3} M.~Ablikim {\em et al.}, [BESIII~Collaboration], Nucl. Instr. Meth. \textbf{A 614}, 345 (2010).\\
               D.~M.~Asner {\em et al.}, [BESIII~Collaboration], Int. J. Mod. Phys. \textbf{A 24}, S1 (2009).

\bibitem{lumi} M.~Ablikim {\em et al.}, [BESIII~Collaboration], Chin. Phys. \textbf{C 37}, 123001 (2013).

\bibitem{TMVA} A.~Hoecker, P.~Speckmayer, J.~Stelzer, J.~Therhaag, E.~Von~Toerne and H.~Voss, PoS ACAT 040 (2007).

\bibitem{phok} G.~Rodrigo, H.~Czy\.{z}, J.~H.~Kuhn, M.~Szopa, Eur. Phys. J. \textbf{C 24}, 71 (2002).\\
               H.~Czyz, J.~H.~Kuhn and A.~Wapienik, Phys. Rev. \textbf{D 77}, 114005 (2008).

\bibitem{radi} V.~Druzhinin, S.~Eidelman, S.~Serednyakov and E.~Solodov, Rev. Mod. Phys. \textbf{83}, 1545 (2011).
               
\bibitem{andi} A.~Hafner, Contribution to these proceedings.               
              

\end{thebibliography}

\begin{thebibliography}{99}
\bibitem{vepp} V.V. Danilov {\em et al.,\/} Proceedings of EPAC96, Barcelona,
p.1593 (1996). \\ I.A. Koop, Nucl. Phys. B (Proc. Suppl.) {\bf 181-182}, 371 
(2008).
\bibitem{sndcmd3} B.I. Khazin,  Nucl. Phys. B (Proc. Suppl.) {\bf 181-182}, 
376 (2008).
\bibitem{dc} F. Grancagnolo {\em et al.,\/} Nucl. Instr. and Meth. A {\bf 623},
114 (2010).
\bibitem{lxe} A.V. Anisyonkov {\em et al.,\/}  Nucl. Instr. and Meth. A 
{\bf 598}, 266 (2009)
\bibitem{cal} D. Epifanov (CMD-3 Collaboration), J. Phys. Conf. Ser. {\bf 293},
012009 (2011).
\bibitem{lum} R.R. Akhmetshin {\em et al.,\/} Nucl. Phys. B (Proc. Suppl.) 
{\bf 225-227}, 69 (2012).
\bibitem{6pi} R.R. Akhmetshin {\em et al.,\/} Phys.Lett. B{\bf 723} 82 (2013).
\bibitem{4pi2pi0} P.A. Lukin {\em et al.,\/} EPJ Web Conf. {\bf 81}  
02010 (2014).
\bibitem{MCWG_Apr14} J.J. van der Bij {\em et al.,\/} arXiv:1406.4639 [hep-ph].
\bibitem{MCWG_Nov14} C.M. Carloni, {\em et al.,\/} arXiv:1412.7714 [hep-ph]. 
\end{thebibliography}

\vspace{-6mm}
\begin{thebibliography}{99}
\vspace{-3mm}

\bibitem{belle2}
 D.Y.~Kim,
To appear in proceedings of the 37th International Conference on High Energy Physics (ICHEP 2014).

\bibitem{belle2b}
 T.~Schlueter,
To appear in proceedings of the 23rd International Workshop on Vertex Detectors (VERTEX 2014).

\bibitem{babayaga}
G.~Balossini at al.,
Nucl. Phys. B758, 227-253, 2006.

\bibitem{bhwide}
S.~Jadach et al.,
Phys. Lett. B390 (1997) 298.

\bibitem{kkmc1}
S.~Jadach et al.,
Comput. Phys. Commun. 130 (2000) 260.

\bibitem{kkmc2}
S.~Jadach et al.,
Phys. Rev. D63 (2001) 113009.

\bibitem{phokhara}
H.~Czyz,
JHEP 1308 (2013) 110.

\bibitem{koralw}
S.~Jadach et al., 
Comput. Phys. Commun. 140 (2001) 475

\bibitem{aafh}
F.~A.~Berends et al., 
Nucl. Phys. 40, B253, 441-463 (1985).

\bibitem{bbrem}
R.~Kleiss et al., Comput.Phys.Commun. 81 (1994) 372-380.

\bibitem{evtgen}
D.~J.~Lange,
Nucl.Instrum.Meth. A462 (2001) 152-155.

\bibitem{tauola}
http://tauolapp.web.cern.ch/tauolapp/

\bibitem{photos}
http://photospp.web.cern.ch/photospp/

\bibitem{madevent}
http://http://madgraph.hep.uiuc.edu/

\bibitem{pythia8}
T. Sjöstrand, S. Mrenna and P. Skands, JHEP05 (2006) 026, Comput. Phys. Comm. 178 (2008) 852.

\end{thebibliography}

\begin{thebibliography}{99}

\bibitem{EMS}
J.~Erler, P.~Masjuan, and H.~Spiesberger, \emph{ in Preparation}.

\bibitem{Novikov:1977dq}
V.~A. Novikov et~al.
\newblock {\em Phys. Rept.}, 41:1--133, 1978.

\bibitem{Shifman:1978bx}
M.~A. Shifman, A.~I. Vainshtein, and V.~I. Zakharov.
\newblock {\em Nucl. Phys.}, B147:385--447, 1979.

\bibitem{Shifman:1978by}
M.~A. Shifman, A.~I. Vainshtein, and V.~I. Zakharov.
\newblock {\em Nucl. Phys.}, B147:448--518, 1979.

\bibitem{Kuhn:2007vp}
J.~H. Kuhn, M.~Steinhauser, and C.~Sturm.
\newblock {\em Nucl. Phys.}, B778:192--215, 2007.

\bibitem{Erler:2002bu}
J.~Erler and M.-X. Luo.
\newblock {\em Phys. Lett.}, B558:125--131, 2003.

\bibitem{Chetyrkin:2000zk}
K.~G. Chetyrkin, R.~V. Harlander, and J.~H. Kuhn.
\newblock {Quartic mass corrections to R(had) at order $\alpha^3(s)$}.
\newblock {\em Nucl. Phys.}, B586:56--72, 2000.

\bibitem{Agashe:2014kda}
K.~A. Olive et~al.
\newblock {\em Chin. Phys.}, C38:090001, 2014.

\bibitem{Kiyo:2009gb}
Y.~Kiyo, A.~Maier, P.~Maierhofer, and P.~Marquard.
\newblock {\em Nucl. Phys.}, B823:269--287, 2009.

\bibitem{Hoang:2008qy}
A.~H. Hoang, V.~Mateu, and S.~Mohammad~Zebarjad.
\newblock {\em Nucl. Phys.}, B813:349--369, 2009.

\bibitem{Greynat:2011zp}
  D.~Greynat, P.~Masjuan and S.~Peris,
  Phys.\ Rev.\ D {\bf 85} (2012) 054008
  [arXiv:1104.3425 [hep-ph]].

\bibitem{Vermaseren:1997fq}
J.~A.~M. Vermaseren, S.~A. Larin, and T.~van Ritbergen.
\newblock {\em Phys. Lett.}, B405:327--333, 1997.

\bibitem{Bai:1999pk}
J.~Z. Bai et~al.
\newblock {\em Phys. Rev. Lett.}, 84:594--597, 2000.

\bibitem{Bai:2001ct}
J.~Z. Bai et~al.
\newblock {\em Phys. Rev. Lett.}, 88:101802, 2002.

\bibitem{Ablikim:2006mb}
M.~Ablikim, J.~Z. Bai, Y.~Ban, J.~G. Bian, X.~Cai, et~al.
\newblock {\em Phys. Rev. Lett.}, 97:262001, 2006.

\bibitem{Ablikim:2006aj}
M.~Ablikim et~al.
\newblock {\em Phys. Lett.}, B641:145--155, 2006.

\bibitem{Ablikim:2009ad}
M.~Ablikim et~al.
\newblock {\em Phys. Lett.}, B677:239--245, 2009.

\bibitem{CroninHennessy:2008yi}
D.~Cronin-Hennessy et~al.
\newblock {\em Phys. Rev.}, D80:072001, 2009.

\end{thebibliography}

\vspace{-3mm}
\begin{thebibliography}{99}
\vspace{-3mm}
\bibitem{carlomat2} K. Ko\l odziej, Comput. Phys. Commun. {\bf 185} (2014) 323, 
      [arXiv:1305.5096].
\bibitem{equiv} G. Ecker, J. Gasser, H. Leutwyler, A. Pich, E. de Rafael, Phys.
                Lett. B 223 (1989) 425.
\bibitem{carlomat} K. Ko\l odziej, Comput. Phys. Commun. {\bf 180} (2009)
1671, [arXiv:0903.3334].
\bibitem{carlomat3} K. Ko\l odziej, {\tt carlomat\_3.0,} an automatic tool
                   for the electron--positron annihilation into hadrons at
                   low energies, Computer Physics Communications (2015),
                   http://dx.doi.org/10.1016/j.cpc.2015.06.013
                   [arXiv:1504.05915 [hep-ph]].
\bibitem{Benayoun} M. Benayoun, P. David, L. DelBuono, F. Jegerlehner,
                    Eur. Phys. J. C 72 (2012) 1848 [arXiv:1106.1315].  
\end{thebibliography}

\begin{thebibliography}{99}
\bibitem{Masjuan:2014rea}
  P.~Masjuan,
  Nucl.\ Part.\ Phys.\ Proc.\  {\bf 260} (2015) 111
  [arXiv:1411.6397 [hep-ph]].
\bibitem{Adlarson:2014hka}
  P.~Adlarson {\it et al.},
  arXiv:1412.5451 [nucl-ex].
  \bibitem{Benayoun:2014tra}
  M.~Benayoun {\it et al.},
  arXiv:1407.4021 [hep-ph].
\bibitem{Pennington}D. Morgan and M.R. Pennington, Z. Phys. {\bf C37} (1988) 431; D. Morgan and M.R. Pennington, Z. Phys. {\bf C48} (1990) 623.
\bibitem{Drechsel:1999rf}
  D.~Drechsel, M.~Gorchtein, B.~Pasquini and M.~Vanderhaeghen,
  Phys.\ Rev.\ C {\bf 61} (1999) 015204
  [hep-ph/9904290].
\bibitem{GarciaMartin:2010cw}
  R.~Garcia-Martin and B.~Moussallam,
  Eur.\ Phys.\ J.\ C {\bf 70} (2010) 155
  [arXiv:1006.5373 [hep-ph]].
\bibitem{Hoferichter:2011wk}
  M.~Hoferichter, D.~R.~Phillips and C.~Schat,
  Eur.\ Phys.\ J.\ C {\bf 71} (2011) 1743
  [arXiv:1106.4147 [hep-ph]].
 \bibitem{Dai:2014zta}
  L.~-Y.~Dai and M.~R.~Pennington,
  arXiv:1404.7524 [hep-ph].
\bibitem{MSPV}
P.~Masjuan, P. Sanchez-Puertas, and M.~Vanderhaeghen, \emph{ in Preparation}.

\bibitem{belle_pic}T. Mori {\it et al.} [Belle], Phys. Rev. {\bf D75} (2007) 051101, {arXiv:0610038 [hep-ex]},
J. Phys. Soc. Jap. {\bf 76} (2007) 074102, {arXiv:0704.3538 [hep-ph]}.
\bibitem{belle_pin}K. Abe {\it et al.} [Belle],  {arXiv:0711.1926 [hep-ex]}; S. Uehara {\it et al.} [Belle], Phys. Rev. {\bf D78} (2008) 052004, {arXiv:0805.3387 [hep-ex]}; S. Uehara {\it et al.} [Belle], Phys. Rev. {\bf D79} (2009) 052009, {arXiv:0903.3697 [hep-ex]}.

\bibitem{Moussallam:2013una}
  B.~Moussallam,
  Eur.\ Phys.\ J.\ C {\bf 73} (2013) 2539
  [arXiv:1305.3143 [hep-ph]].

\bibitem{GarciaMartin:2011cn}
  R.~Garcia-Martin, R.~Kaminski, J.~R.~Pelaez, J.~Ruiz de Elvira and F.~J.~Yndurain,
  Phys.\ Rev.\ D {\bf 83} (2011) 074004
  [arXiv:1102.2183 [hep-ph]].

\bibitem{Masjuan:2012gc}
  P.~Masjuan, E.~Ruiz Arriola and W.~Broniowski,
  Phys.\ Rev.\ D {\bf 85} (2012) 094006
  [arXiv:1203.4782 [hep-ph]].

\bibitem{Masjuan:2008fv} 
  P.~Masjuan, S.~Peris and J.~J.~Sanz-Cillero,
  Phys.\ Rev.\ D {\bf 78}, 074028 (2008)
  [arXiv:0807.4893 [hep-ph]].

\bibitem{Masjuan:2012sk}
  P.~Masjuan, E.~Ruiz Arriola and W.~Broniowski,
  Phys.\ Rev.\ D {\bf 87} (2013) 1,  014005
  [arXiv:1210.0760 [hep-ph]].
\bibitem{Escribano:2013kba}
  R.~Escribano, P.~Masjuan and P.~Sanchez-Puertas,
  Phys.\ Rev.\ D {\bf 89} (2014) 034014
  [arXiv:1307.2061 [hep-ph]];
 


\end{thebibliography}

\begin{thebibliography}{99} 
\vspace{-3mm} 
\bibitem{Jadach:1993hs}
  S.~Jadach, Z.~Was, R.~Decker and J.~H.~Kuhn,
  Comput.\ Phys.\ Commun.\  {\bf 76} (1993) 361.

\bibitem{Buskulic:1995ty}
  D.~Buskulic {\it et al.}  [ALEPH Collaboration],
  Z.\ Phys.\ C {\bf 70} (1996) 579.

\bibitem{Asner:1999kj}
  D.~M.~Asner {\it et al.}  [CLEO Collaboration],
  Phys.\ Rev.\ D {\bf 61} (2000) 012002
  [hep-ex/9902022].

\bibitem{Nugent:2013ij}
  I.~M.~Nugent [BaBar Collaboration],
  Nucl.\ Phys.\ Proc.\ Suppl.\  {\bf 253-255} (2014) 38
  [arXiv:1301.7105 [hep-ex]].

\bibitem{Fujikawa:2008ma}
  M.~Fujikawa {\it et al.}  [Belle Collaboration],
  Phys.\ Rev.\ D {\bf 78} (2008) 072006
  [arXiv:0805.3773 [hep-ex]].


\bibitem{Aad:2012tfa}
  G.~Aad {\it et al.}  [ATLAS Collaboration],
  Phys.\ Lett.\ B {\bf 716} (2012) 1
  [arXiv:1207.7214 [hep-ex]].

\bibitem{Abe:2010gxa}
  T.~Abe {\it et al.}  [Belle-II Collaboration],
  arXiv:1011.0352 [physics.ins-det].

\bibitem{Kuhn:1990ad}
  J.~H.~Kuhn and A.~Santamaria,
  Z.\ Phys.\ C {\bf 48} (1990) 445.

\bibitem{SanzCillero:2002bs}
  J.~J.~Sanz-Cillero and A.~Pich,
  Eur.\ Phys.\ J.\ C {\bf 27} (2003) 587
  [hep-ph/0208199].

\bibitem{Dumm:2013zh}
  D.~Gómez Dumm and P.~Roig,
  Eur.\ Phys.\ J.\ C {\bf 73} (2013) 8,  2528
  [arXiv:1301.6973 [hep-ph]].

\bibitem{Shekhovtsova:2012ra}
  O.~Shekhovtsova, T.~Przedzinski, P.~Roig and Z.~Was,
  Phys.\ Rev.\ D {\bf 86} (2012) 113008
  [arXiv:1203.3955 [hep-ph]].

\bibitem{Shibata:2002uv}
  E.~I.~Shibata [CLEO Collaboration],
  eConf C {\bf 0209101} (2002) TU05
   [Nucl.\ Phys.\ Proc.\ Suppl.\  {\bf 123} (2003) 40]
  [hep-ex/0210039].


\bibitem{Nugent:2013hxa}
  I.~M.~Nugent, T.~Przedzinski, P.~Roig, O.~Shekhovtsova and Z.~Was,
  Phys.\ Rev.\ D {\bf 88} (2013) 9,  093012
  [arXiv:1310.1053 [hep-ph]].


\end{thebibliography}

\begin{thebibliography}{99}

  \bibitem{Bennett:2006fi}
  G.W.~Bennett {\it et al.}  [Muon g-2 Collaboration],
  Phys.\ Rev.\ D {\bf 73} (2006) 072003.

  \bibitem{Jegerlehner:2009ry}
  F.~Jegerlehner, A.~Nyffeler,
  Phys.\ Rept.\  {\bf 477} (2009) 1.

  \bibitem{Reviews}
  T.~Blum {\it et al.}, 
  arXiv:1311.2198 [hep-ph];
   K.~Melnikov, A.~Vainshtein,
  Springer Tracts Mod.\ Phys.\  {\bf 216} (2006) 1;
   M.~Davier, W.J.~Marciano,
  Ann.\ Rev.\ Nucl.\ Part.\ Sci.\  {\bf 54} (2004) 115;
   M.~Passera,
  J.\ Phys.\ G {\bf 31} (2005) R75;
   M.~Knecht,
  Lect.\ Notes Phys.\  {\bf 629} (2004) 37.
   
   \bibitem{Jegerlehner:2008zza} 
   F.~Jegerlehner,
   ``The anomalous magnetic moment of the muon,''
   Springer Tracts Mod.\ Phys.\  {\bf 226}, 2008.

  \bibitem{Grange:2015fou}
  J.~Grange {\it et al.}  [Muon g-2 Collaboration],
  arXiv:1501.06858 [physics.ins-det].

\bibitem{Venanzoni:2014ixa}
  G.~Venanzoni [Muon g-2 Collaboration],
  arXiv:1411.2555 [physics.ins-det].

\bibitem{Saito:2012zz}
  N.~Saito [J-PARC g-2/EDM Collaboration],
  AIP Conf.\ Proc.\  {\bf 1467} (2012) 45.
  
\bibitem{Venanzoni:2014wva}
  G.~Venanzoni,
  Nuovo Cim.\ C {\bf 037} (2014) 02,  165; 
  G.~Venanzoni,
  Frascati Phys.\ Ser.\  {\bf 54} (2012) 52.
    
\bibitem{Fedotovich:2008zz}
  G.V.~Fedotovich [CMD-2 Collaboration],
  Nucl.\ Phys.\ Proc.\ Suppl.\  {\bf 181-182} (2008) 146.

  \bibitem{Krause:1996rf}
  B.~Krause,
  Phys.\ Lett.\ B {\bf 390} (1997) 392.
  
  \bibitem{HLBL}
  M.~Knecht, A.~Nyffeler,
  Phys.\ Rev.\ D {\bf 65} (2002) 073034;
  K.~Melnikov, A.~Vainshtein,
  Phys.\ Rev.\ D {\bf 70} (2004) 113006;
  J.~Prades, E.~de Rafael, A.~Vainshtein,
  arXiv:0901.0306 [hep-ph].

  \bibitem{HLBLfuture}
  G.~Colangelo, M.~Hoferichter, M.~Procura, P.~Stoffer,
  JHEP {\bf 1409} (2014) 091;
  G.~Colangelo, M.~Hoferichter, B.~Kubis, M.~Procura, P.~Stoffer,
  Phys.\ Lett.\ B {\bf 738} (2014) 6;
  V.~Pauk, M.~Vanderhaeghen,
  Phys.\ Rev.\ D {\bf 90} (2014) 11,  113012;
  T.~Blum, S.~Chowdhury, M.~Hayakawa, T.~Izubuchi,
  Phys.\ Rev.\ Lett.\  {\bf 114} (2015) 1,  012001.
    
  \bibitem{Kurz:2014wya}
  A.~Kurz, T.~Liu, P.~Marquard, M.~Steinhauser,
  Phys.\ Lett.\ B {\bf 734} (2014) 144.
   
  \bibitem{Colangelo:2014qya}
  G.~Colangelo, M.~Hoferichter, A.~Nyffeler, M.~Passera, P.~Stoffer,
  Phys.\ Lett.\ B {\bf 735} (2014) 90.

 \bibitem{BM61GDR69}  
  C.~Bouchiat, L.~Michel, J.~Phys.~Radium 22 (1961) 121;
  L.~Durand,
  Phys.\ Rev.\  {\bf 128} (1962) 441 [Erratum-ibid.\ {\bf 129} (1963) 2835];
  M.~Gourdin, E.~De Rafael,
  Nucl.\ Phys.\ B {\bf 10} (1969) 667.
 
\bibitem{Calame:2015fva}
  C.~M.~Carloni Calame, M.~Passera, L.~Trentadue and G.~Venanzoni,
  Phys.\ Lett.\ B {\bf 746} (2015) 325
  [arXiv:1504.02228 [hep-ph]].
 \end{thebibliography}
\end{document}